\documentclass{article}
\usepackage{ijcai26}

\usepackage{times}
\usepackage{soul}
\usepackage{url}
\usepackage[hidelinks]{hyperref}
\usepackage[utf8]{inputenc}
\usepackage[small]{caption}
\usepackage{graphicx}
\usepackage{amsmath}
\usepackage{amsthm}
\usepackage{booktabs}
\usepackage{algorithm}
\usepackage{algorithmic}
\usepackage[switch]{lineno}

\usepackage{physics}
\usepackage{makecell}
\usepackage{amsfonts}
\usepackage{xcolor}
\usepackage{placeins}

\title{End-to-End QGAN-Based Image Synthesis via Neural Noise Encoding and Intensity Calibration}

\author{
    \mdseries 
    Xue Yang$^{1,2}$ \quad
    Rigui Zhou$^{1}$\setcounter{footnote}{1}\thanks{Corresponding authors.} \quad
    Shizheng Jia$^{1}$ \quad
    Dax Enshan Koh$^{2, 3\dagger}$ \\
    Siong Thye Goh$^{4\dagger}$ \quad
    Yaochong Li$^{1}$ \quad
    Hongyu Chen$^{1}$ \quad
    Fuhui Xiong$^{1}$ \\[2mm]
    $^1$Shanghai Maritime University \quad
    $^2$Agency for Science, Technology and Research \\
    $^3$Singapore University of Technology and Design \\
    $^4$Institute of High Performance Computing \\[2mm]
}

\begin{document}

\maketitle

\begin{abstract}
Quantum Generative Adversarial Networks (QGANs) offer a promising path for learning data distributions on near-term quantum devices. However, existing QGANs for image synthesis avoid direct full-image generation, relying on classical post-processing or patch-based methods. These approaches dilute the quantum generator's role and struggle to capture global image semantics. To address this, we propose ReQGAN, an end-to-end framework that synthesizes an entire $N=2^D$-pixel image using a single $D$-qubit quantum circuit. ReQGAN overcomes two fundamental bottlenecks hindering direct pixel generation: (1) the rigid classical-to-quantum noise interface and (2) the output mismatch between normalized quantum statistics and the desired pixel-intensity space. We introduce a learnable Neural Noise Encoder for adaptive state preparation and a differentiable Intensity Calibration module to map measurements to a stable, visually meaningful pixel domain. Experiments on MNIST and Fashion-MNIST demonstrate that ReQGAN achieves stable training and effective image synthesis under stringent qubit budgets, with ablation studies verifying the contribution of each component.
\end{abstract}


\section{Introduction}

\begin{figure}[!htbp]
    \centering
    \includegraphics[width=\linewidth]{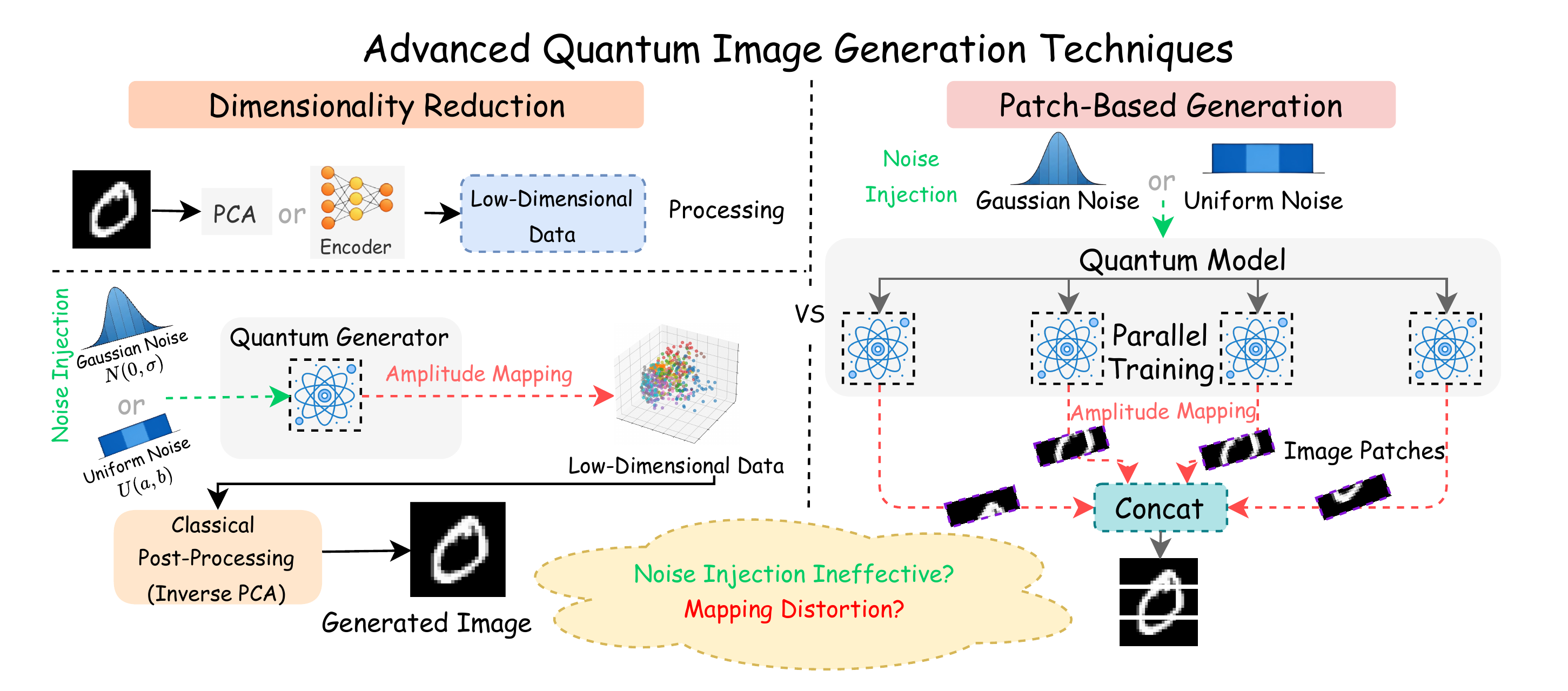}
    \caption{Common strategies for QGAN-based image synthesis. Left: Low-dimensional generation relies on classical components for dimensionality reduction and reconstruction. Right: Patch-wise generation uses multiple quantum sub-generators for local patches. Both approaches circumvent two key bottlenecks for direct end-to-end generation, which we highlight: ineffective noise injection and the challenge of mapping normalized quantum outputs to pixel intensities.}
    \label{fig:intro}
\end{figure}

Generative modeling is a central task in modern machine learning, and Generative Adversarial Networks (GANs) have become a canonical framework for learning high-dimensional data distributions via adversarial training \cite{goodfellow2014gan}.
In parallel, quantum machine learning has explored whether quantum devices can offer new representational or computational capabilities for learning problems \cite{biamonte2017qml}.
Quantum Generative Adversarial Networks (QGANs) bring these lines together by instantiating the generator and/or discriminator with quantum models \cite{lloyd2018qgal,dallaire2018qgan}, and have been demonstrated in early proof-of-principle experiments \cite{hu2019qgan}.
A key motivation is that an $n$-qubit quantum state lives in a $2^n$-dimensional Hilbert space, enabling compact representations of high-dimensional vectors through amplitude-style encodings \cite{nielsen2010qcqi,schuld2019featurehilbert}.
While preparing arbitrary amplitudes can be resource-intensive in general \cite{grover2002stateprep,soklakov2006stateprep}, the exponential size of the state space still makes QGANs an appealing candidate for modeling complex distributions with limited physical resources.

Despite this promise, near-term Noisy Intermediate-Scale Quantum (NISQ) devices impose tight constraints on qubit count and circuit depth \cite{preskill2018nisq,cerezo2021vqa}, and training parameterized quantum circuits (PQCs) can suffer from optimization pathologies such as barren plateaus \cite{mcclean2018barrenplateaus}.
Consequently, most QGAN studies targeting image synthesis focus on MNIST-scale benchmarks and adopt engineering strategies that avoid direct, full-resolution generation.
As summarized in Fig.~\ref{fig:intro}, existing methods largely follow two routes.
Low-dimensional generation compresses images using classical dimensionality reduction (e.g., principal component analysis (PCA) \cite{jolliffe2002pca} or autoencoder-style encoders), allowing the quantum generator to model only the compact representation, and then relies on a classical decoder to reconstruct the image.
This approach is often practical, but it tightly couples success to classical pre-/post-processing, which can obscure the real contribution of the quantum generator.
Patch-wise generation instead targets pixel space directly by training multiple quantum sub-generators to synthesize local patches that are later stitched into a full image \cite{huang2021expqgan,tsang2022hybrid}.
While effective under severe qubit limits, patch-wise schemes may struggle to capture global structure and tend to scale quantum resources with the number of patches.

These observations raise a natural question: why is end-to-end, pixel-level full-image generation from a single quantum state still uncommon in QGANs?
In this paper, we argue that two mechanism-level mismatches make naive ``one-shot'' QGAN image synthesis hard to train.
First, the latent noise interface is often implemented by directly feeding classical noise components into fixed single-qubit rotations \cite{huang2021expqgan,tsang2022hybrid}, which is convenient but may be poorly aligned with the expressive characteristics of a given PQC.
More broadly, data/noise encoding strategies can dominate the behavior of variational quantum models \cite{schuld2021dataencoding,perezsalinas2020datareuploading}.
Second, the quantum-to-image mapping is intrinsically constrained by quantum normalization: measurement probabilities (or amplitudes) live on a simplex (or a unit sphere), whereas natural images require flexible global intensity and contrast statistics.
Simple probability-to-pixel mappings (e.g., max-normalization) can therefore collapse magnitude information and destabilize training, motivating a principled, differentiable calibration from quantum statistics to pixel intensities.

To address these challenges, we propose ReQGAN, an end-to-end QGAN framework that synthesizes an $N$-pixel image using a single PQC with $D=\log_2 N$ data qubits.
ReQGAN introduces two coordinated components.
(1) A Neural Noise Encoder adaptively transforms raw latent noise into circuit-driving parameters, learning a noise injection scheme matched to the PQC and simultaneously predicting sample-conditioned affine coefficients for downstream calibration.
(2) An Intensity Calibration module converts the measured conditional distribution into a stable pixel domain through a structured cascade (smoothing, deviation-driven modeling, contrast normalization, and adaptive affine projection), explicitly correcting the scale/normalization mismatch between quantum outputs and images.
Together, these designs enable end-to-end adversarial training for full-image pixel synthesis under stringent qubit budgets.

\paragraph{Contributions.}
\begin{itemize}
    \item We propose ReQGAN, an end-to-end QGAN framework that performs full-image pixel generation with a single PQC, avoiding patch stitching and explicit classical reconstruction pipelines.
    \item We introduce a learnable Neural Noise Encoder to replace rigid noise-to-circuit heuristics with an adaptive, trainable interface.
    \item We design a structured Intensity Calibration module that robustly maps normalized quantum measurement statistics into a visually coherent pixel-intensity space.
    \item We empirically validate ReQGAN on MNIST and Fashion-MNIST and provide ablation analyses that isolate the impact of the Neural Noise Encoder and Intensity Calibration on training stability and synthesis quality.
\end{itemize}

\section{Related Work and Background}

\paragraph{Foundations of Adversarial Learning.}
GANs learn a data distribution via a minimax game between a generator $G$ and a discriminator $D$ \cite{goodfellow2014gan}. To enhance training stability, variants like Wasserstein GAN (WGAN) and WGAN-GP introduce the Wasserstein-1 distance and a gradient penalty, respectively \cite{arjovsky2017wgan,gulrajani2017wgangp}. This adversarial training paradigm has been extended to quantum settings, where the generator and/or discriminator can be quantum models \cite{lloyd2018qgal,dallaire2018qgan}, with early experiments demonstrating feasibility on contemporary hardware \cite{hu2019qgan}. 

\paragraph{Prevalent Routes in QGAN Image Synthesis.}
Recent QGAN research for image synthesis has increasingly targeted MNIST-scale benchmarks to move beyond toy problems \cite{tsang2022hybrid,silver2023mosaiq,chu2022iqgan,YANG2026128865}. However, current approaches predominantly follow two engineering routes that circumvent direct, full-image generation. (i) Patch-wise generation composes a full image by stitching together outputs from multiple smaller quantum sub-generators \cite{huang2021expqgan,tsang2022hybrid}. (ii) Low-dimensional generation first down-samples high-dimensional data (e.g., via PCA), generates new samples in this compact latent space, and then reconstructs them back to the full image size \cite{chu2022iqgan,silver2023mosaiq}. While these strategies have enabled practical progress, they also indicate that one-shot, end-to-end generation of a full image from a single quantum state remains a less explored area.

\paragraph{The Latent Noise Interface Challenge.}
A critical design choice in QGANs is the interface mapping classical latent noise to the quantum circuit. A common practice is to use latent vector entries as parameters for fixed single-qubit rotations (e.g., $RY(\alpha z(i))$) \cite{huang2021expqgan,tsang2022hybrid}. This heuristic approach is convenient but not necessarily optimal. Several recent studies argue this interface is a key factor for performance. For instance, IQGAN found an adjustable input encoder crucial for quality \cite{chu2022iqgan}, MosaiQ introduced adaptive input noise generation to improve variety \cite{silver2023mosaiq}, and it is known more generally that the data-encoding strategy strongly impacts the expressive power of variational models \cite{schuld2021dataencoding}. These findings motivate treating the noise-to-circuit mapping as a learnable component.

\paragraph{The Quantum-to-Image Mapping and Normalization Challenge.}
The output mapping from a quantum state to an image presents another significant challenge. While various quantum image representations exist \cite{le2011frqi,zhang2013neqr}, QGANs often decode pixel values from measurement probabilities or amplitudes \cite{huang2021expqgan}. A core issue with these approaches stems from the inherent normalization in quantum mechanics. Formally, in an amplitude-style encoding, a real vector $x \in \mathbb{R}^{2^n}$ is mapped to a quantum state where the amplitudes are normalized:
$|x\rangle = \frac{1}{\|x\|_2}\sum_{j=0}^{2^n-1} x_j |j\rangle$.
This forced normalization, explicit in the $1/\|x\|_2$ term, can collapse magnitude information and alter the global intensity and contrast semantics of an image, which are crucial for visual fidelity \cite{kiwit2025lowdepth,morgan2026qrnn}. Consequently, the inherently normalized quantum output space is not naturally aligned with the target image space, making the quantum-to-image map a first-class modeling problem.

\section{The Proposed Framework}

\begin{figure*}[htb]
    \centering
    \includegraphics[width=\textwidth]{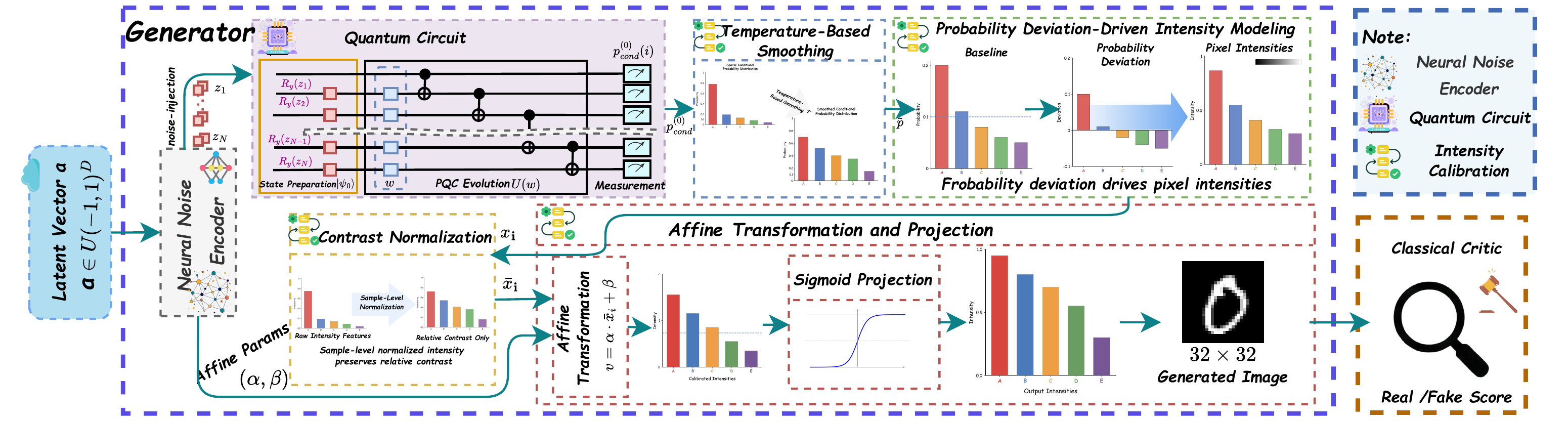}
    \caption{Schematic diagram of the overall ReQGAN framework. The generator consists of a Neural Noise Encoder, a Quantum Circuit, and an Intensity Calibration module. The generator first samples a latent vector $\mathbf{a}$ from a uniform distribution and inputs it into the Neural Noise Encoder. This encoder outputs two types of information: a noise vector $\mathbf{z}$ injected into the quantum circuit to drive quantum state preparation, and affine calibration coefficients $(\alpha, \beta)$ used for subsequent intensity calibration. The conditional probability distribution measured from the quantum circuit then enters the Intensity Calibration module, sequentially undergoing amplitude smoothing, deviation modeling, contrast normalization, and adaptive affine projection, to be converted into pixel intensity representations. Finally, the generator and discriminator are alternately updated and jointly optimized under an adversarial training framework to achieve end-to-end pixel-level image generation.}
    \label{fig:overview}
\end{figure*}

\subsection{Overview}
We propose ReQGAN, an end-to-end quantum generative adversarial framework that synthesizes an $N$-pixel image using a single PQC, where $N = 2^D$ and $D$ denotes the number of data qubits. As illustrated in Fig.~\ref{fig:overview}, the generator is composed of three sequential modules: a Neural Noise Encoder, a Quantum Circuit, and an Intensity Calibration module. The generation process begins by sampling a classical noise vector $\mathbf{a} \sim U(-1,1)^D$, which is fed into the Neural Noise Encoder. This encoder outputs two components: (i) a noise-injection vector $\mathbf{z} \in \mathbb{R}^D$ to control the quantum state preparation, and (ii) an affine parameter pair $\boldsymbol{\phi}=(\alpha,\beta)$ for adaptive pixel intensity calibration. The quantum circuit then evolves the prepared state, and a subsequent measurement in the computational basis yields a probability distribution. Critically, instead of directly interpreting this distribution as pixel values, our Intensity Calibration module applies a differentiable cascade of transformations—smoothing, deviation-driven modeling, contrast normalization, and affine projection—to robustly map the quantum statistics into a valid and visually coherent pixel domain.

\subsection{Neural Noise Encoder}
The Neural Noise Encoder serves as the interface between classical stochasticity and the quantum generator. Given $\mathbf{a} \sim U(-1,1)^D$, the encoder $f_\theta(\cdot)$ produces a noise-injection vector $\mathbf{z} \in \mathbb{R}^D$ and affine parameters $\boldsymbol{\phi}=(\alpha,\beta)$:
\begin{equation}
    (\mathbf{z}, \boldsymbol{\phi}) = f_\theta(\mathbf{a}).
\end{equation}
This dual-output design is intentional. The vector $\mathbf{z}$ modulates the initial quantum state, diversifying the probability landscapes produced by the PQC. Concurrently, the parameters $(\alpha,\beta)$ provide a learnable, sample-conditioned mechanism to restore global brightness and contrast information that might be lost during subsequent normalization steps, ensuring stable and consistent image synthesis.

\subsection{Quantum Circuit}
\begin{figure}
    \centering
\includegraphics[width=0.5\textwidth]{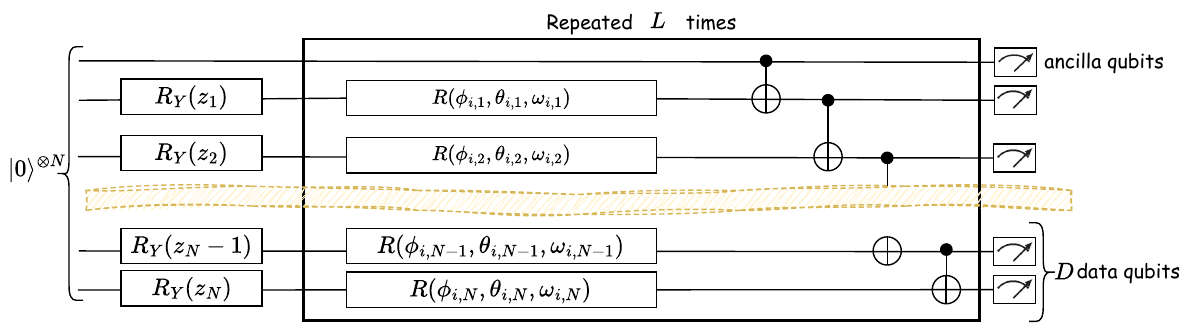}
    \caption{Schematic of the quantum generator circuit consisting of state preparation $P(\mathbf{z})$ and an $L$-layer parameterized ansatz.}
    \label{fig:circuit}
\end{figure}

The core of our generator is a quantum circuit, shown in Fig.~\ref{fig:circuit}, which acts on $D$ data qubits and one ancilla qubit. It consists of a state preparation layer $P(\mathbf{z})$ followed by an $L$-layer PQC $U(\boldsymbol{\omega})$.

\paragraph{State Preparation.}
The system is initialized in the all-zero state $\ket{0}^{\otimes(D+1)}$. The noise-injection vector $\mathbf{z}$ from the encoder then prepares the data register by applying controlled $y$-axis rotations to each data qubit, creating the initial state:
\begin{equation}
\ket{\psi_0}
=
\left(\bigotimes_{i=1}^{D} R_y(z_i)\right)\ket{0}^{\otimes D}\otimes\ket{0}_{\mathrm{anc}}.
\end{equation}

\paragraph{PQC Evolution.}
Following preparation, the state $\ket{\psi_0}$ is evolved by the $L$-layer PQC, which alternates between trainable single-qubit rotations and fixed entangling operations (CNOT gates). Denoting the $l$-th layer as $U_l(\boldsymbol{\omega}_l)$ with trainable parameters $\boldsymbol{\omega}_l$, the full unitary is given by:
\begin{equation}
U(\boldsymbol{\omega}) = U_L(\boldsymbol{\omega}_L)\cdots U_2(\boldsymbol{\omega}_2)U_1(\boldsymbol{\omega}_1),
\end{equation}
where $\boldsymbol{\omega}=\{\boldsymbol{\omega}_1,\dots,\boldsymbol{\omega}_L\}$. The final state of the system is thus $\ket{\psi} = U(\boldsymbol{\omega})\ket{\psi_0}$.

\paragraph{Measurement and Conditional Output.}
A measurement of the final state $\ket{\psi}$ in the computational basis yields a joint probability distribution $p(\mathbf{i}, b) = \left|\braket{\mathbf{i}, b}{\psi}\right|^2$, where $\mathbf{i}\in\{0,1\}^D$ is the outcome of the data qubits and $b\in\{0,1\}$ is the ancilla outcome. We then derive a conditional distribution over the data space by post-selecting on the ancilla being measured as $0$:
\begin{equation}
p_{\mathrm{cond}}^{(0)}(\mathbf{i})
=
\frac{p(\mathbf{i},0)}{\sum_{\mathbf{j}\in\{0,1\}^D} p(\mathbf{j},0)}.
\end{equation}
This conditional distribution $p_{\mathrm{cond}}^{(0)}$ serves as the raw quantum output. We interpret each basis string $\mathbf{i}$ as a pixel location (via its integer representation), yielding an $N=2^D$ dimensional vector that is then processed by the Intensity Calibration module.

\subsection{Intensity Calibration}
Conventional QGANs often directly map measurement probabilities to pixel intensities, a process prone to instability. We introduce a robust, fully differentiable calibration module that transforms the raw distribution $p_{\mathrm{cond}}^{(0)}$ into a meaningful image. This is achieved through a four-stage cascade.

\paragraph{Temperature-Based Smoothing.}
The raw conditional distribution $p_{\mathrm{cond}}^{(0)}$ can be highly peaked, concentrating learning signals on a few basis states. To mitigate this and encourage a broader feature representation, we apply temperature-based smoothing:
\begin{equation}
\tilde{p}_{\mathbf{i}}
=
\frac{\left(p_{\mathrm{cond}}^{(0)}(\mathbf{i})+\epsilon_p\right)^{1/\tau}}
{\sum_{\mathbf{j}}\left(p_{\mathrm{cond}}^{(0)}(\mathbf{j})+\epsilon_p\right)^{1/\tau}},
\qquad \tau>1,
\end{equation}
where $\tau$ is a temperature hyperparameter that flattens the distribution, and $\epsilon_p > 0$ ensures numerical stability.

\paragraph{Probability Deviation-Driven Intensity Modeling.}
We reframe the smoothed distribution $\tilde{\mathbf{p}}$ as evidence of relative saliency rather than absolute intensity. To quantify this, we measure its deviation from a non-informative baseline, the uniform distribution $p^{\mathrm{uni}}_{\mathbf{i}}=1/N$. This relative deviation is defined as:
\begin{equation}
u_{\mathbf{i}}
= N\tilde{p}_{\mathbf{i}} - 1
= \frac{\tilde{p}_{\mathbf{i}} - 1/N}{1/N}.
\end{equation}
Here, $u_{\mathbf{i}}>0$ signifies higher-than-average probability concentration. To amplify these salient signals while suppressing negative (less-than-average) interference, we apply a non-linear mapping:
\begin{equation}
x_{\mathbf{i}}
=
\mathrm{softplus}(k u_{\mathbf{i}}) - \mathrm{softplus}(0),
\qquad k>0,
\end{equation}
which ensures $x_{\mathbf{i}}=0$ when $\tilde{p}_{\mathbf{i}}$ matches the uniform baseline ($u_{\mathbf{i}}=0$).

\paragraph{Contrast Normalization.}
The vector $\mathbf{x}=(x_{\mathbf{i}})_{\mathbf{i}}$ captures spatial structure but may exhibit sample-wise variations in global scale and offset. To remove this drift and focus the model on learning relative patterns, we apply instance-level contrast normalization. For each sample, we compute its mean $\mu$ and standard deviation $\sigma$:
\begin{equation}
\mu = \frac{1}{N}\sum_{\mathbf{i}} x_{\mathbf{i}},
\qquad
\sigma = \sqrt{\frac{1}{N}\sum_{\mathbf{i}}(x_{\mathbf{i}}-\mu)^2},
\end{equation}
and normalize each component:
\begin{equation}
\bar{x}_{\mathbf{i}} = \frac{x_{\mathbf{i}}-\mu}{\sigma+\epsilon_n},
\end{equation}
where $\epsilon_n>0$ is a small stabilizer. This step standardizes the dynamic range of each generated sample.

\paragraph{Affine Transformation and Projection.}
Contrast normalization discards absolute intensity information. To reintroduce controllable brightness and contrast in a sample-specific manner, we apply a final affine transformation using the parameters $(\alpha, \beta)$ predicted by the Neural Noise Encoder:
\begin{equation}
v_{\mathbf{i}} = \alpha \bar{x}_{\mathbf{i}} + \beta.
\end{equation}
Finally, the resulting values are projected into a valid pixel range using an element-wise function $\pi(\cdot)$. We use the sigmoid function, $\pi(\cdot) = \sigma(\cdot)$, to map intensities to the $(0,1)$ range:
\begin{equation}
x^{\mathrm{img}}_{\mathbf{i}} = \pi(v_{\mathbf{i}}).
\end{equation}
The resulting vector $\mathbf{x}^{\mathrm{img}}$ can be reshaped into the final image.

\subsection{Adversarial Training}
The ReQGAN framework is trained end-to-end within a standard adversarial setting. For a fair comparison with prior work and to avoid introducing additional inductive biases, we employ a purely classical discriminator with an identical architecture to that used in PQWGAN. 
The training proceeds via a min-max game, where we alternate between updating the discriminator and the generator. The generator's loss is backpropagated through the entire differentiable pipeline, jointly optimizing the classical encoder parameters $\theta$ and the quantum circuit parameters $\boldsymbol{\omega}$. 

\section{Experiments}
\subsection{Datasets}
Due to the limited qubit resources of near-term NISQ hardware, most QGAN studies validate on standard grayscale image benchmarks. Following this established practice and to enable fair comparison with prior work, we use MNIST \cite{lecun1998mnist} and Fashion-MNIST \cite{xiao2017fashionmnist}, both consisting of 10 classes of 28$\times$28 grayscale images. We used 1,250 images from each dataset and split them into training and test sets at an 8:2 ratio, resulting in 1,000 training images and 250 test images.

\subsection{Implementation Details}
All experiments were implemented in TensorFlow \cite{abadi2016tensorflow_osdi} with TensorCircuit \cite{zhang2023tensorcircuit_quantum} for quantum circuit simulation. Numerical simulations were conducted on a server with an AMD EPYC 9654 CPU and 128 GB RAM. We used the Adam optimizer for all trainable components. The initial learning rate was set to $2\times10^{-4}$ for the classical discriminator and the neural noise encoder, whereas the PQC was optimized with a higher initial learning rate of $10^{-2}$. The Adam momentum parameters were fixed to $\beta_1=0.0$ and $\beta_2=0.9$. All models were trained for 50 epochs with a batch size of 5.

The objective of this study is to achieve direct pixel-level synthesis of full images in an end-to-end manner using a single quantum circuit. Accordingly, we select \textit{PQWGAN} \cite{tsang2022hybrid} as the baseline for comparison. As a representative of existing QGAN design paradigms, PQWGAN without patch-wise decomposition (i.e., \textit{Patch} $=1$) effectively attempts a naive end-to-end image synthesis. By directly comparing our method with PQWGAN under this setting, we can explicitly evaluate the effectiveness of the proposed architecture in overcoming the inherent limitations of conventional approaches.

\subsection{Evaluation Metrics}
To quantitatively assess both the quality and diversity of the generated images, we adopt two widely established metrics: Fr\'echet Inception Distance (FID) \cite{heusel2017ttur_fid} and Kernel Inception Distance (KID) \cite{binkowski2018kid}. FID measures the Fr\'echet distance between feature distributions of real and generated images, with lower FID values indicating superior quality and better diversity:
\begin{equation}
\small
\text{FID}(P_r, P_g) = \|\mu_r - \mu_g\|^2 + \text{Tr}\left(\Sigma_r + \Sigma_g - 2(\Sigma_r \Sigma_g)^{1/2}\right)
\end{equation}
where $  (\mu_r, \Sigma_r)  $ and $  (\mu_g, \Sigma_g)  $ denote the mean and covariance of the feature vectors extracted from real and generated images, respectively. And KID computes the squared MMD with a polynomial kernel in the same feature space, providing an unbiased estimate suitable for small sample sizes:
\begin{equation}
\small
\begin{aligned}
\text{KID}(P_r,& P_g) = \text{MMD}^2(P_r, P_g) = \mathbb{E}_{x,x'\sim P_r}[k(x,x')] \\&+ \mathbb{E}_{y,y'\sim P_g}[k(y,y')] - 2\mathbb{E}_{x\sim P_r,y\sim P_g}[k(x,y)]
\end{aligned}
\end{equation}
where $  k(\cdot,\cdot)  $ is the polynomial kernel function used to measure similarity in the feature space. A lower KID value indicates that the generated distribution is closer to the real data distribution.

In addition, to assess whether quantum measurement outcomes are properly mapped to a reasonable pixel-intensity scale, we report average brightness and RMS contrast as intensity statistics. In addition, we monitor the Wasserstein distance to quantify the convergence behavior of model.


\section{Results}

\subsection{Comparison with the Baseline Model}
We evaluate ReQGAN against PQWGAN in terms of training behavior and generation quality using both quantitative metrics and qualitative comparisons.

\begin{figure*}[t]
    \centering
    \includegraphics[width=0.85\linewidth]{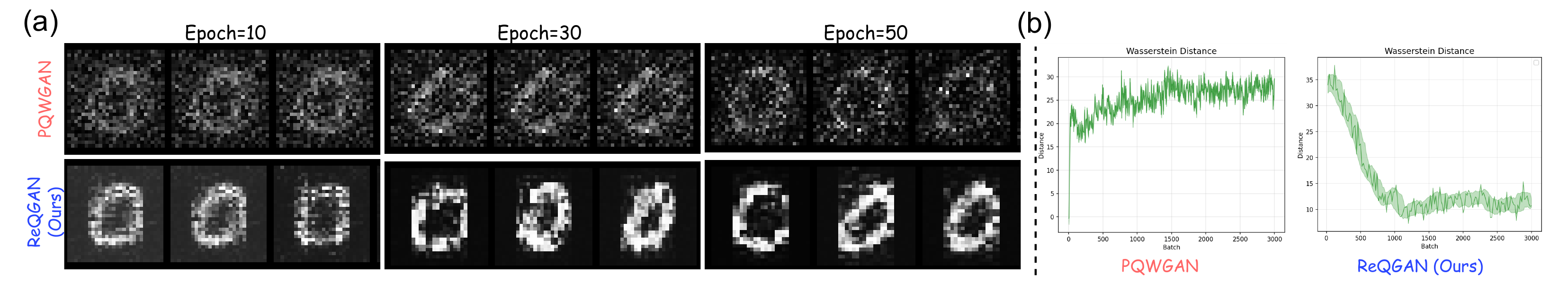}
    \caption{Visualization of real samples and generated/perturbed fake samples (top row), and the corresponding average FFT spectra before (middle row) and after (bottom row) the Edge-to-RGB completion.}
    \label{fig:visual}
\end{figure*}

\paragraph{Training stability and qualitative comparison.}
We conduct a controlled comparison on the MNIST single-class subset of digit 0. Figure~\ref{fig:visual} provides representative visual comparisons between real samples and generated samples (including perturbed variants), together with their corresponding average FFT spectra before and after the Edge-to-RGB completion. Overall, ReQGAN produces cleaner digit structures and backgrounds, while PQWGAN is more prone to unstable artifacts during optimization. In addition, we monitor the Wasserstein objective during training and observe that ReQGAN converges more smoothly, whereas PQWGAN exhibits larger fluctuations.

\begin{figure}[!htbp]
    \centering
    \includegraphics[width=0.85\linewidth]{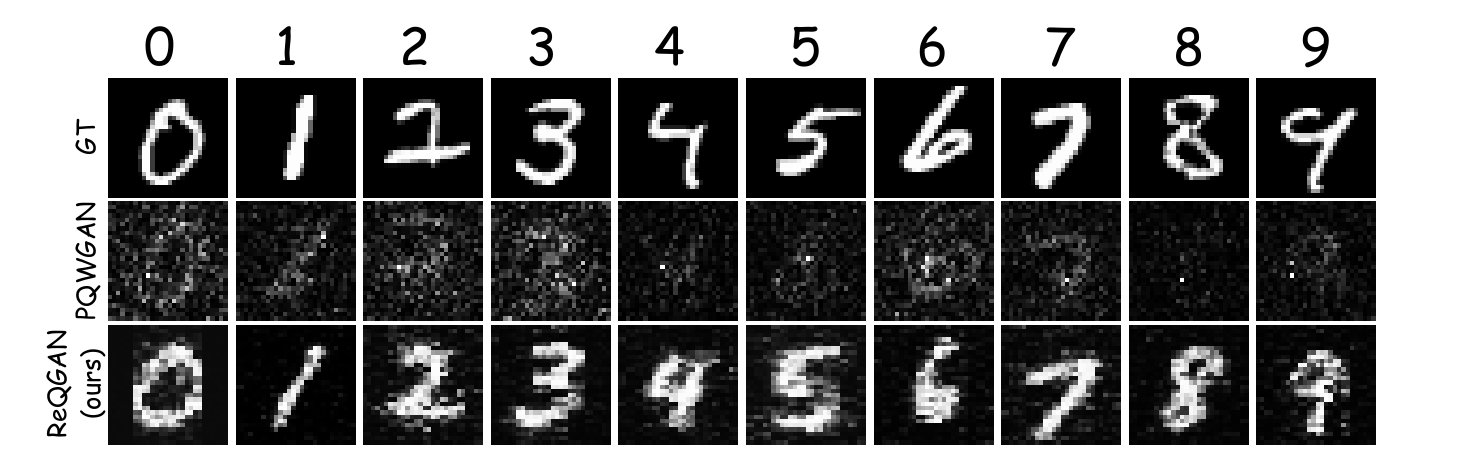}
    \caption{Generated samples of PQWGAN and ReQGAN across MNIST digit classes (0--9).}
    \label{fig:digit}
\end{figure}

\paragraph{Performance across MNIST digit classes.}
We further compare the two methods across all MNIST digit classes (0--9). Each class is trained independently, and the generated samples are evaluated using FID and KID. Figure~\ref{fig:digit} shows qualitative results across classes, where ReQGAN generally yields clearer contours and fewer background artifacts. Table~\ref{tab:mnist_vertical} reports class-wise FID/KID scores, showing that ReQGAN achieves lower FID and KID on most digit classes, indicating improved robustness across categories.

\begin{table}[!htbp]
\renewcommand\arraystretch{1.05}
\centering
\scriptsize
\begin{tabular}{c|cc|cc}
\toprule
& \multicolumn{2}{c|}{PQWGAN} & \multicolumn{2}{c}{ReQGAN (ours)} \\
\cmidrule(lr){2-3} \cmidrule(lr){4-5}
& FID$\downarrow$ & KID$\downarrow$ & FID$\downarrow$ & KID$\downarrow$ \\
\midrule
Digit 0 & 89.46 $\pm$ 0.35 & 0.1931 $\pm$ 0.0039 & \textbf{45.53} $\pm$ 0.11 & \textbf{0.0436} $\pm$ 0.0023 \\
Digit 1 & 42.11 $\pm$ 0.11 & 0.0834 $\pm$ 0.0006 & \textbf{12.54} $\pm$ 0.11 & \textbf{0.0092} $\pm$ 0.0010 \\
Digit 2 & 79.88 $\pm$ 0.09 & 0.1347 $\pm$ 0.0047 & \textbf{44.25} $\pm$ 0.07 & \textbf{0.0241} $\pm$ 0.0019 \\
Digit 3 & 73.02 $\pm$ 0.09 & 0.1361 $\pm$ 0.0031 & \textbf{40.27} $\pm$ 0.07 & \textbf{0.0259} $\pm$ 0.0011 \\
Digit 4 & 68.44 $\pm$ 0.12 & 0.1237 $\pm$ 0.0022 & \textbf{32.83} $\pm$ 0.07 & \textbf{0.0187} $\pm$ 0.0002 \\
Digit 5 & 67.82 $\pm$ 0.08 & 0.0994 $\pm$ 0.0037 & \textbf{37.54} $\pm$ 0.30 & \textbf{0.0203} $\pm$ 0.0011 \\
Digit 6 & 68.54 $\pm$ 0.06 & 0.1376 $\pm$ 0.0034 & \textbf{30.82} $\pm$ 0.12 & \textbf{0.0204} $\pm$ 0.0006 \\
Digit 7 & 57.10 $\pm$ 0.10 & 0.0931 $\pm$ 0.0026 & \textbf{27.49} $\pm$ 0.29 & \textbf{0.0240} $\pm$ 0.0025 \\
Digit 8 & 87.81 $\pm$ 0.16 & 0.1932 $\pm$ 0.0077 & \textbf{33.93} $\pm$ 0.29 & \textbf{0.0192} $\pm$ 0.0028 \\
Digit 9 & 67.81 $\pm$ 0.03 & 0.1370 $\pm$ 0.0031 & \textbf{36.25} $\pm$ 0.09 & \textbf{0.0277} $\pm$ 0.0016 \\
\bottomrule
\end{tabular}
\caption{Class-wise comparison on MNIST (digits 0--9). Lower is better for both FID and KID. The best result in each row is highlighted in bold.}
\label{tab:mnist_vertical}
\end{table}

\subsection{Ablation Experiments of ReQGAN}
To assess the contribution of each component in ReQGAN, we conduct ablation studies under a single-class MNIST setting. For each ablation group, the digit class is fixed across variants to ensure a fair comparison.

\begin{figure}[t]
    \centering
    \includegraphics[width=0.8\linewidth]{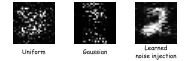}
    \caption{Qualitative results for the noise injection ablation in Table~\ref{tab:ablation_noise_injection}.}
    \label{fig:tab2}
\end{figure}

\paragraph{Noise injection in the quantum circuit.}
We study the impact of the learned noise injection mechanism on the MNIST single-class subset of digit 2. When the learned noise injection is removed and replaced by i.i.d.\ noise sampled from $U(0,1)$ or $\mathcal{N}(0,1)$, generation quality degrades substantially. Table~\ref{tab:ablation_noise_injection} reports the quantitative results, and Figure~\ref{fig:tab2} shows qualitative comparisons. The learned noise injection provides effective regularization and improves both stability and visual fidelity.

\begin{table}[t]
\centering
\renewcommand\arraystretch{1.05}
\footnotesize
\setlength{\tabcolsep}{3.5pt}
\begin{tabular}{c|ccc}
\toprule
& \multicolumn{2}{c|}{w/o learned noise injection} & ReQGAN (ours) \\
\cmidrule(lr){2-3}\cmidrule(lr){4-4}
& \begin{tabular}[c]{@{}c@{}}Uniform\\ $U(0,1)$\end{tabular}
& \begin{tabular}[c]{@{}c@{}}Gaussian\\ $\mathcal{N}(0,1)$\end{tabular}
& \begin{tabular}[c]{@{}c@{}}Learned\\ noise injection\end{tabular} \\
\midrule
FID$\downarrow$ & 62.23 $\pm$ 0.09 & 78.39 $\pm$ 0.80 & 44.25 $\pm$ 0.07 \\
KID$\downarrow$ & 0.0662 $\pm$ 0.0027 & 0.1397 $\pm$ 0.0036 & 0.0241 $\pm$ 0.0019 \\
\bottomrule
\end{tabular}
\caption{Ablation on the noise injection mechanism in the quantum circuit (MNIST digit 2). Lower is better for both FID and KID.}
\label{tab:ablation_noise_injection}
\end{table}

\begin{figure}[!htbp]
    \centering
    \includegraphics[width=\linewidth]{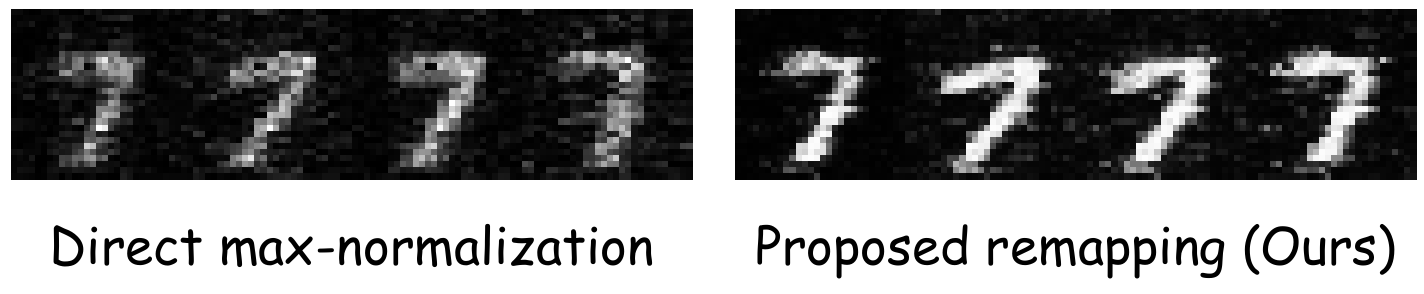}
    \caption{Qualitative results for the mapping ablation in Table~\ref{tab:ablation_mapping_reQgan}.}
    \label{fig:tab3}
\end{figure}

\paragraph{Probability-to-intensity mapping.}
We evaluate whether the probability-to-intensity mapping affects generation quality on the MNIST single-class subset of digit 7. We compare the proposed intensity calibration mapping in ReQGAN against a direct max-normalization baseline (``/max'' mapping). Table~\ref{tab:ablation_mapping_reQgan} shows that the proposed mapping improves FID/KID and yields more appropriate brightness/contrast statistics. Figure~\ref{fig:tab3} provides qualitative comparisons: the baseline mapping tends to produce samples with lower brightness and insufficient contrast, while ReQGAN produces cleaner backgrounds and more visually consistent digits.

\begin{table}[t]
\centering
\renewcommand\arraystretch{1.05}
\footnotesize
\setlength{\tabcolsep}{2pt}
\begin{tabular}{c|cc}
\toprule
& Baseline mapping & ReQGAN (ours) \\
\midrule
FID$\downarrow$ & 38.78 $\pm$ 0.12 & 27.49 $\pm$ 0.29 \\
KID$\downarrow$ & 0.0272 $\pm$ 0.0021 & 0.0240 $\pm$ 0.0025 \\
Avg Brightness$\uparrow$ & 29.6261 $\pm$ 2.6922 & 31.7062 $\pm$ 9.3662 \\
RMS Contrast$\uparrow$ & 43.8794 $\pm$ 4.0255 & 61.4014 $\pm$ 8.0891 \\
\bottomrule
\end{tabular}
\caption{Ablation on the probability-to-intensity mapping strategy (MNIST digit 7). The baseline rescales the circuit output by its maximum value (``/max'' mapping), while ReQGAN uses the proposed intensity calibration. Lower is better for FID/KID; higher is better for brightness and contrast.}
\label{tab:ablation_mapping_reQgan}
\end{table}

\begin{table*}[!h]
\centering
\renewcommand\arraystretch{1}
\small
\setlength{\tabcolsep}{2pt}
\resizebox{1\textwidth}{!}{
\begin{tabular}{l|cccccccccc}
\toprule
& class 0 & class 1 & class 2 & class 3 & class 4 & class 5 & class 6 & class 7 & class 8 & class 9 \\
\midrule
FID$\downarrow$ & 52.21 $\pm$ 0.36 & 37.15 $\pm$ 0.08 & 54.92 $\pm$ 0.67 & 48.97 $\pm$ 0.28 & 45.21 $\pm$ 0.27 & 82.81 $\pm$ 0.34 & 63.33 $\pm$ 0.67 & 51.93 $\pm$ 0.38 & 73.67 $\pm$ 0.79 & 62.69 $\pm$ 0.02 \\
KID$\downarrow$ & 0.1415 $\pm$ 0.0027 & 0.0973 $\pm$ 0.0040 & 0.1242 $\pm$ 0.0038 & 0.1027 $\pm$ 0.0028 & 0.0868 $\pm$ 0.0004 & 0.2125 $\pm$ 0.0041 & 0.1643 $\pm$ 0.0105 & 0.1653 $\pm$ 0.0034 & 0.1427 $\pm$ 0.0028 & 0.1587 $\pm$ 0.0083 \\
\bottomrule
\end{tabular}
}
\caption{Class-wise FID and KID scores of ReQGAN on FashionMNIST. Lower is better for both FID and KID.}
\label{tab:fashion_reqgan_A}
\end{table*}

\begin{figure}[!htbp]
    \centering
    \includegraphics[width=\linewidth]{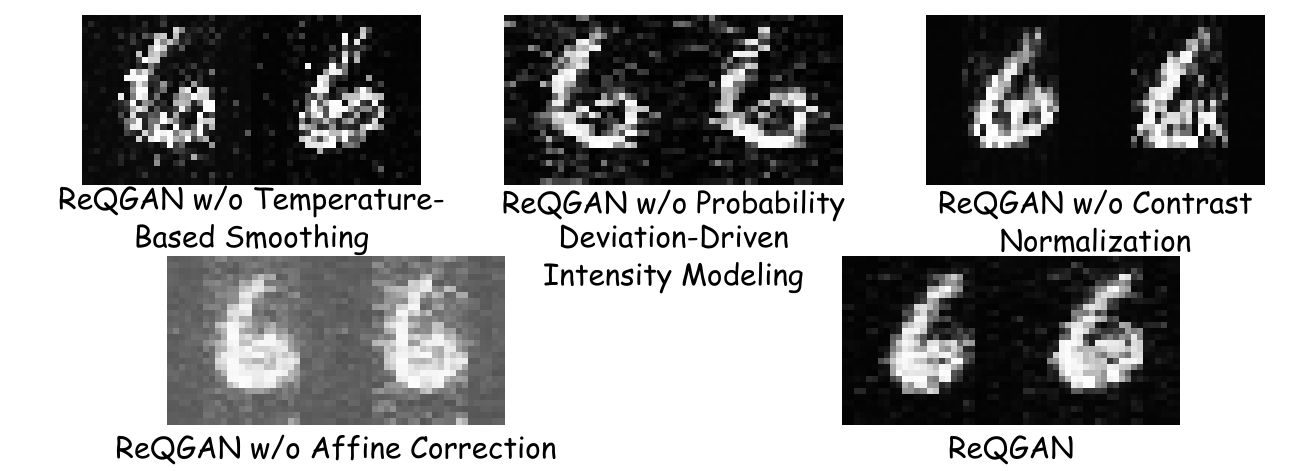}
    \caption{Qualitative results for the submodule ablation in Table~\ref{tab:ablation_revgan_left_breaks}.}
    \label{fig:tab4}
\end{figure}

\paragraph{Ablation of individual intensity calibration submodules.}
We analyze the roles of the submodules within the intensity calibration module on the MNIST single-class subset of digit 6. We remove four submodules individually and report the quantitative results in Table~\ref{tab:ablation_revgan_left_breaks}, with qualitative comparisons in Figure~\ref{fig:tab4}. Removing temperature-based smoothing yields overly peaked probability patterns and increases discrete noise in pixel space. Removing deviation-driven intensity modeling makes it harder to suppress background activity, resulting in fragmented contours. Without contrast normalization, overall brightness becomes less consistent across samples, although degradation is moderate. Removing affine correction causes the most severe degradation, introducing global brightness bias and reduced contrast. Overall, all submodules contribute positively, and affine correction is particularly important for stable and high-fidelity generation.

\begin{table}[!htbp]
\centering
\renewcommand\arraystretch{1.05}
\scriptsize
\setlength{\tabcolsep}{4pt}
\begin{tabular}{p{0.54\linewidth}|cc}
\toprule
Method & FID$\downarrow$ & KID$\downarrow$ \\
\midrule
\makecell[l]{ReQGAN w/o Temperature-Based Smoothing} & 33.88 $\pm$ 0.12 & 0.0219 $\pm$ 0.0012 \\
\makecell[l]{ReQGAN w/o Probability Deviation-Driven \\ Intensity Modeling} & 33.42 $\pm$ 0.07 & 0.0223 $\pm$ 0.0013 \\
\makecell[l]{ReQGAN w/o Contrast Normalization} & 33.97 $\pm$ 0.32 & 0.0167 $\pm$ 0.0008 \\
\makecell[l]{ReQGAN w/o Affine Correction} & 142.31 $\pm$ 0.02 & 0.6229 $\pm$ 0.0056 \\
\makecell[l]{ReQGAN} & 30.82 $\pm$ 0.12 & 0.0204 $\pm$ 0.0006 \\
\bottomrule
\end{tabular}
\caption{Ablation study on submodules of the intensity calibration module in ReQGAN (MNIST digit 6). Lower is better for both FID and KID.}
\label{tab:ablation_revgan_left_breaks}
\end{table}

\subsection{Generalization Performance of ReQGAN}
\begin{figure}[!htbp]
    \centering
    \includegraphics[width=\linewidth]{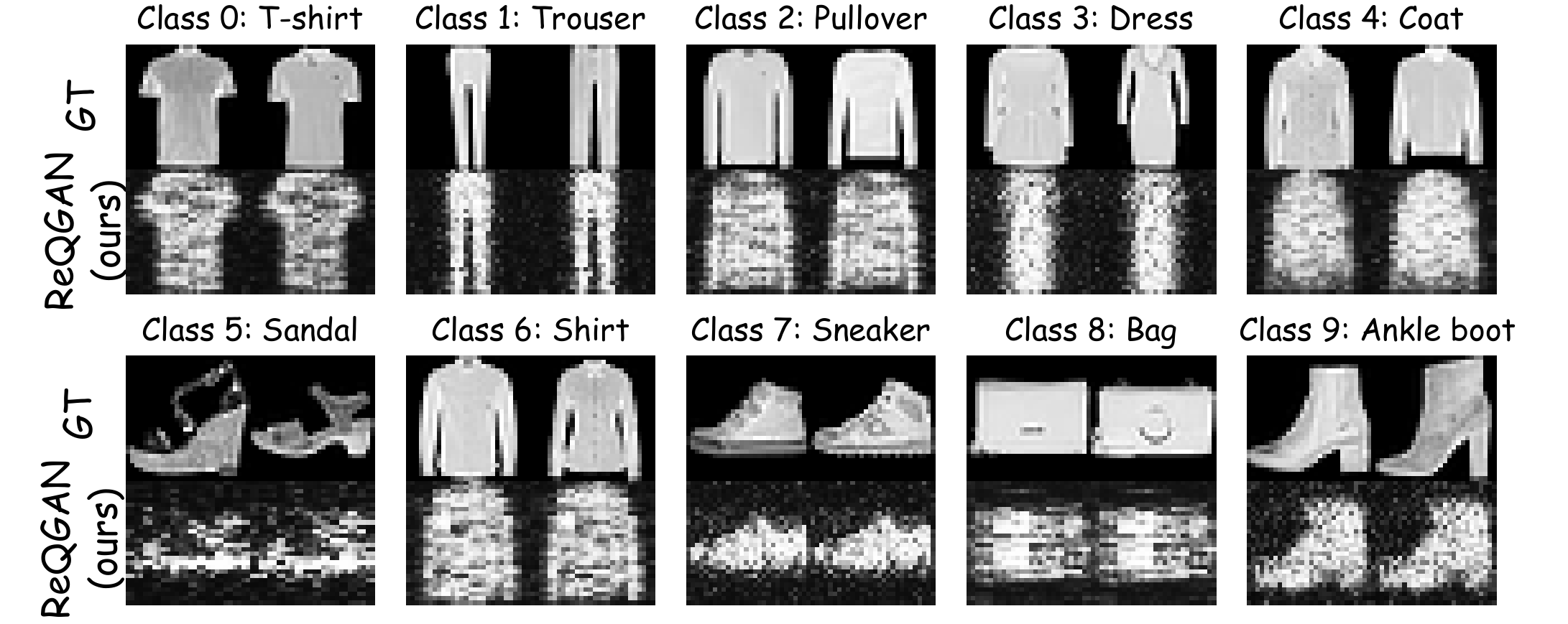}
    \caption{Generated samples of ReQGAN across FashionMNIST classes (0--9).}
    \label{fig:Fashion}
\end{figure}

To evaluate generalization, we train ReQGAN on FashionMNIST and generate samples across all ten categories (0--9) without changing the architecture or training configuration. Figure~\ref{fig:Fashion} shows that ReQGAN produces structurally coherent and visually distinguishable samples across classes. The class-wise quantitative results are reported in Table~\ref{tab:fashion_reqgan_A}, indicating stable generative performance under a different data distribution.

\section{Conclusion}
This paper studied the bottlenecks that limit end-to-end, pixel-level image synthesis in QGANs under NISQ constraints.
We argued that two mismatches are particularly critical: a rigid classical-to-quantum noise interface and the normalization-induced misalignment between quantum outputs and image intensity statistics.
To overcome these issues, we presented ReQGAN, which synthesizes a full $N=2^D$-pixel image from a single PQC by combining a learnable Neural Noise Encoder with a differentiable Intensity Calibration module.
Across MNIST-scale benchmarks, ReQGAN enables stable end-to-end adversarial training and effective image synthesis within strict qubit budgets, while ablations confirm that both the Neural Noise Encoder and Intensity Calibration are key to convergence and quality.
Looking forward, promising directions include extending calibration to color images and higher resolutions, incorporating shot-noise-aware training for hardware execution, and exploring more expressive yet trainable circuit ans\"atze and encoding strategies.

\FloatBarrier

\bibliographystyle{named}
\bibliography{ijcai26} 

\end{document}